\documentclass[12pt, prd, showpacs]{revtex4}
\usepackage{amssymb}
\usepackage{amsmath}

\begin{document}

\title{Redshift of a photon emitted along the black hole horizon}
\author{A. \ V. Toporensky}
\affiliation{Sternberg Astronomical Institute, Lomonosov Moscow State University }
\affiliation{Kazan Federal University, Kremlevskaya 18, Kazan 420008, Russia}
\email{atopor@rambler.ru}
\author{O. B. Zaslavskii}
\affiliation{Department of Physics and Technology, Kharkov V.N. Karazin National
University, 4 Svoboda Square, Kharkov 61022, Ukraine}
\affiliation{Kazan Federal University, Kremlevskaya 18, Kazan 420008 Russia}
\email{zaslav@ukr.net}

\begin{abstract}
In this work we derive some general features of the redshift measured by
radially moving observers in the black hole background. Let observer 1 crosses
the black hole horizon emitting a photon while observer 2 crossing the
same horizon later receives it. We show that if (i) the horizon is the outer
one (event horizon) and (ii) it is nonextremal, the received frequency is
redshifted. This generalizes previous recent results in literature. For the
inner horizon (like in the Reissner-Nordstr\"{o}m metric) the frequency is
blueshifted. If the horizon is extremal, the frequency does not change. We
derive explicit formulas describing the frequency shift in generalized
Kruskal- and Lemaitre-like coordinates.
\end{abstract}

\maketitle

\section{Introduction}

Redshift is one of well-known effects of gravity, it plays an essential
role in relativistic astrophysics. Its description entered many textbooks.
In particles, it concerns propagation of light in the black hole background.
Thus main attention is focused on properties of light outside the
event horizon. Less  these properties inside the horizon are discussed.
Also, strange as it may seem, the question of the redshift of a photon
moving along the horizon dropped out from consideration almost completely.
There is general discussion of this issue in \cite{lake} for the
Reissner-Nordstr\"{o}m - de Sitter metric but the relation between redshift or
blueshift and the nature of the horizon was not revealed there.

Meanwhile, there are several points that can serve as a motivation for such a
consideration.

(i) Recently, important methodic issues were discussed in \cite{au}, \cite%
{kas} concerning properties of the world visible by an observer falling into
the Schwarzschild black hole and communicating by radio signals with another
falling one. In doing so, some incorrect statements were made in \cite{au}
about "ghosts" of the first observer supposedly waiting for the second one
on the horizon. As was argued in \cite{kas}, there are no such ghosts at all
although the second observer does receive a signal from the first one
emitted at the moment of crossing the horizon. This required detailed
calculation of the frequency shift for a photon propagating along the
horizon with the result that a finite redshift occurs in this case.

In the present work, we generalize these observations, demonstrate that such
a redshift is present for any horizon of a spherically symmetric nonextremal
black hole and find its value.

(ii) If the metric contains an inner horizon (say, for the Reissner-Nordstr%
\"{o}m black hole), the calculation of the frequency
shift for a photon emitted along such a horizon is of special interest. We demonstrate that now,
instead of a redshift, a blueshift occurs.  In the limit when a photon
is received near the bifurcation point, the blueshift becomes infinite. This
establishes the connection of the issue under discussion with the analogue
of the Ba\~{n}ados-Silk-West (BSW) effect \cite{ban}. It consists in the
infinite growth of the energy of colliding particles in the centre of mass
frame. Originally, it was found near the event horizon but, later on, it
turned out that another similar effect is valid also near the inner horizon
(see \cite{inner} and references therein). From another hand, the issue
under discussion can be considered as an effect supplemental to a well-known
instability of the inner horizon \cite{fn}, \cite{ham}.

(iii) In addition to the propagation along the nonextremal horizon, there is
also a question what happens in the extremal case. We argue that, by
contrast to two previous ones, now the frequency shift is absent.

(iv) In papers \cite{au}, \cite{kas} the Kruskal-Szekerez (KS) coordinate
system was used. We also exploit it. In addition, it is of interest to
compare the results using another powerful system - the Lemaitre one. We
construct such a system for a whole class of metrics that includes the
Schwarzchild one as a particular case.

In the present work, we restrict ourselves by the simplest case of radially
moving observers.

\section{Motion outside the nonextremal event horizon}

We consider the metric%
\begin{equation}
ds^{2}=-fdt^{2}+\frac{dr^{2}}{f}+r^{2}(d\theta ^{2}+\sin ^{2}\theta d\phi
^{2})\text{.}  \label{met}
\end{equation}

We suppose that the metric has the event horizon at $r=r_{+}$, so $%
f(r_{+})=0 $. (For simplicity, we assumed that $g_{00}g_{11}=-1$ but this
condition can be relaxed easily.) We consider now a nonextremal black hole.
Near the horizon,%
\begin{equation}
f\approx \kappa (r-r_{+})\text{,}  \label{kar}
\end{equation}%
where $\kappa =\frac{f^{\prime }(r_{+})}{2}$ is the surface gravity.

Let an observer has the four-velocity $u^{\mu }=\frac{dx^{\mu }}{d\tau }$,
where $\tau $ is the proper time. We restrict ourselves to the radial motion
of a massive particle (we call it "observer"). Then, the four-velocity $%
u^{\mu }=(\dot{t}$, $\dot{r})$. Here, a dot denotes derivative with respect to 
$\tau $.

The geodesic equations of motion for such a particle read:%
\begin{equation}
m\dot{t}=\frac{E}{f}\text{,}  \label{mt}
\end{equation}%
\begin{equation}
m\dot{r}=-Z\text{, }Z=\sqrt{E^{2}-fm^{2}}\text{,}  \label{mr}
\end{equation}%
$E=-mu_t$ is the conserved energy  of a particle.

For a photon having the wave vector $k^{\mu }$, the equations of motion are%
\begin{equation}
k^{t}=\frac{\omega _{0}}{f}\text{,}  \label{yt}
\end{equation}%
\begin{equation}
k^{\phi }=\frac{l}{g_{\phi }}\text{,}
\end{equation}%
\begin{equation}
k^{r}=-Q\text{, }Q=\sqrt{\omega _{0}^{2}-f\frac{l^{2}}{r^{2}}}  \label{r}
\end{equation}%
where $\omega _{0}=-k_{0}$ is the conserved frequency, $l=k_{\phi }$ is the
conserved angular momentum.

Let a free falling observer emits or receives a photon. Its frequency measured
by this observer is equal to 
\begin{equation}
\omega =-k_{\mu }u^{\mu }.  \label{om}
\end{equation}

Taking into account (\ref{mt}) - (\ref{r}), we find after straightforward
calculation:%
\begin{equation}
\omega =\frac{E\omega _{0}-\varepsilon ZQ}{mf}\text{.}  \label{wzq}
\end{equation}

Here, $\varepsilon =+1$ if both objects (the observer and the photon) move in
the same direction and $\varepsilon =-1$ if they do this in opposite ones.
If $\varepsilon =-1$, this corresponds to head-on collision between a
massive and a massless particles and this means that either an observer receives
a photon emitted from a smaller value of $r$ or a falling observer emits a
photon in the backward direction.

Near the horizon, $f\ll 1.$ Let $\varepsilon =+1$ and the observer looks
back. Near the horizon, he will see the frequency%
\begin{equation}
\omega \approx \frac{1}{2}(\frac{El^{2}}{m\omega _{0}r_{+}^{2}}+\frac{\omega
_{0}m}{E})\text{.}  \label{n1}
\end{equation}%
If $\varepsilon =-1$,%
\begin{equation}
\frac{\omega }{\omega _{0}}\approx \frac{2E}{mf_{em}}\text{.}  \label{n2}
\end{equation}%
Here $f_{em}$ is the value of the metric function in the point of emission.
Let the observer emits in his frame a photon having the frequency $\omega $
and travelling to infinity. Then, at infinity it will be received with the
frequency $\omega _{0}$ having the order $f_{em}$. Eq. (\ref{n2}) agrees
with the standard result for the Schwarzschild metric (see Sec. XII.102 of 
\cite{LL}, especially eq. 102.10).

It is instructive to compare this to another situation usually discussed in
textbooks when the emitter is not free falling but is static. In the latter
case, the frequency at infinity $\omega _{0}=\omega \sqrt{f}$ (see, e.g. eq.
88.6 of \cite{LL}) has the order $\sqrt{f_{em}}\,$. Obviously, the
difference can be attributed to the motion of an emitter (the Doppler effect
makes the redshift stronger).

Some reservations are in order. Throughout the paper, we assume that the
geometric optics is a reasonable approximation for propagation of light
waves. As usual, this implies that the wavelength $\lambda $ satisfies
relations $\frac{\lambda }{2\pi }\ll \mathcal{L},$ $\mathcal{R},$ where $%
\lambda $ is the wave length, $\mathcal{L}$ is a typical scale
characterizing a wave packet and $\mathcal{R}^{-2}$ is the typical component
of the Riemann tensor (see eq. 22.23c in \cite{grav}). For the Schwarzschild
metric, this entails, in terms of the frequency, the condition $\omega M\gg
1,M$ is the black hole mass. In the more general case of, say, the
Reissner-Nordstr\"{o}m metric, this gives $\omega r_{+}\gg 1$, where $r_{+}$
is the radius of the horison.

Also, we assume that backreaction of the photon on the metric is negligible.
This implies that, in any case, its energy is much less than the ADM mass of
a black hole, so in natural geometric units $\omega \ll M$. Thus we have the
double inequality $r_{+}^{-1}\ll \omega \ll M$. As for the Reissner-Nordstr\"{o}m
 metric$\ M\leq r_{+}\leq 2M$, the conditions of validity of this
approximation are practically the same for the Reissner-Nordstr\"{o}m and
Schwarzschild metrics.

\section{Photon emitted at the horizon}

We see from (\ref{n2}) that, as the point of emission of a photon is
approaching the horizon, the frequency measured at infinity becomes smaller
and smaller. However, this formula does not describe what happens if the
photon is emitted exactly on the horizon. Then, $\omega _{0}=0$ and $l=0$, $%
f_{em}=0$ from the very beginning (see also below), the photon does not
reach infinity at all and moves along the leg of the horizon. Here, original
coordinates (\ref{met}) are not applicable since the metric becomes
degenerate on the horizon. This can be remedied with the use of standard KS
coordinates or its generalization.

Let us introduce coordinates $U$ and $V$, where%
\begin{equation}
U=-\exp (-\kappa u)\text{, }V=\exp (\kappa v)\text{,}  \label{uv}
\end{equation}%
\begin{equation}
u=t-r^{\ast }\text{, }v=t+r^{\ast }\text{.}  \label{uU}
\end{equation}%
\begin{equation}
r^{\ast }=\int^{r}\frac{dr}{f}  \label{tc}
\end{equation}%
is the tortoise coordinate. It is seen from (\ref{uv}) that%
\begin{equation}
UV=-\exp (2\kappa r^{\ast })\text{.}  \label{uvr}
\end{equation}

Then the metric reads%
\begin{equation}
ds^{2}=FdUdV+r^{2}d\omega ^{2}\text{,}  \label{mF}
\end{equation}%
Here,%
\begin{equation}
F=-f\frac{du}{dU}\frac{dv}{dV}\text{.}  \label{fg}
\end{equation}%
For the transformation (\ref{uv}),%
\begin{equation}
F=\frac{f}{UV\kappa ^{2}}\text{,}
\end{equation}%
$F\neq 0$ on the horizon. It is clear from (\ref{met}), (\ref{uvr}) that $%
F=F(r)$. For example, for the Schwarzschild metric,%
\begin{equation}
F=-\frac{4r_{+}^{3}}{r}e^{-\frac{r}{r_{+}}}\text{,}
\end{equation}%
provided the constant of integration is chosen in (\ref{tc}) in such a way
that $r^{\ast }=r+r_{+}\ln \frac{r-r_{+}}{r_{+}}$.

We consider the vicinity of the future horizon, on which $U=0$. Along this
horizon, $V$ takes finite values.

Near the horizon%
\begin{equation}
r^{\ast }\approx \frac{1}{2\kappa }\ln \frac{r-r_{+}}{r_{+}}+const\text{,}
\end{equation}%
\begin{equation}
UV\approx \frac{(r-r_{+})}{r_{+}}C_{0}\text{,}  \label{uvc}
\end{equation}%
where $C_{0}$ is a constant. In the Schwarzschild case, $C_{0}=e$. It is
instructive to rewrite equations of motion for massive particles outside the
event horizon (\ref{mt}) - (\ref{mr})\ in terms of the KS coordinates and
take the horizon limit afterwards. For an observer moving inward they read%
\begin{equation}
mu^{U}=\frac{E+Z}{f}\frac{dU}{du}\text{,}  \label{mu}
\end{equation}%
\begin{equation}
mu^{V}=\frac{E-Z}{f}\frac{dV}{dv}\text{.}  \label{mV}
\end{equation}%
Taking also into account (\ref{uv}) and (17), we have%
\begin{equation}
m\dot{U}=-\frac{(E+Z)}{FV\kappa }\text{,}  \label{uu}
\end{equation}%
\begin{equation}
m\dot{V}=\frac{(E-Z)}{UF\kappa }.  \label{vv}
\end{equation}

In a similar way, we have for a photon moving in the outward direction:%
\begin{equation}
k^{U}=\frac{\kappa U}{f}(Q-\omega _{0})=\frac{Q-\omega _{0}}{F\kappa V}\text{%
,}  \label{ku}
\end{equation}%
\begin{equation}
k^{V}=\frac{\kappa V}{f}(\omega _{0}+Q)=\frac{\omega _{0}+Q}{FU\kappa }\text{%
.}
\end{equation}

Let $\lambda $ be the affine parameter. On the future horizon $U=0$, $k^{U}=%
\frac{dU}{d\lambda }=0=k_{V}$ which agrees with (\ref{ku}) if we put $\omega
_{0}=0=f=Q$ in the right hand side. Thus only $k^{V\text{ }}$remains
nonzero. It follows from the geodesic equations that on the horizon%
\begin{equation}
\frac{dk^{V}}{d\lambda }=-\Gamma _{VV}^{V}\left( k^{V}\right) ^{2}\text{.}
\end{equation}%
Here the Christoffel symbol $\Gamma _{VV}^{V}\sim \frac{\partial F}{\partial
V}\sim \frac{\partial F}{\partial r}\frac{\partial r}{\partial V}\sim f=0$
on the horizon. Therefore, $k^{V\text{ }}=const$ along the horizon
generator. We have from (\ref{om}), (\ref{uu}) on the horizon%
\begin{equation}
\omega =-\frac{1}{2} Fk^{V}u^{U}\text{,}  \label{omf0}
\end{equation}%
and we have

\begin{equation}
\omega = \frac{E}{Vm\kappa }k^{V}\text{.}  \label{omf}
\end{equation}

Let observer 1 crosses the horizon at some $V=V_{1}$ and the same for observer 2 but
later, at $V=V_{2}$. It follows from (\ref{uv}), (\ref{uU}) that $V_{2}>V_{1}$%
. Assuming that observers are identical in that they have the same values of 
$E$ and $m$, we obtain

\begin{equation}
\frac{\omega _{2}}{\omega _{1}}=\frac{V_{1}}{V_{2}}<1\text{.}  \label{12}
\end{equation}

This agrees with eq. (A16) of \cite{kas} obtained for the Schwarzschild
metric by another method.

It is also instructive to check that indeed $\omega _{0}=0$. By definition, $%
\omega _{0}$ is a constant Killing frequency 
\begin{equation}
\omega _{0}=-k_{\mu }\xi ^{\mu }\text{,}  \label{fr}
\end{equation}%
where $\xi ^{\mu }$ is the Killing vector. In the original coordinates (\ref%
{met}), 
\begin{equation}
\xi ^{\mu }=(1,0,0,0)\text{, }\xi _{\mu }=(-f,0,0,0)\text{.}
\end{equation}%
Passing to KS coordinates, one obtains%
\begin{equation}
\xi ^{U}=-\kappa U\text{, }\xi ^{V}=\kappa V\text{.}
\end{equation}%
Then, we see from (\ref{fr}) that

\begin{equation}
\omega _{0}=-F(k^{V}\xi ^{U}+k^{U}\xi ^{V})=F\kappa (k^{V}U-k^{U}V)\text{.}
\end{equation}%
On the future horizon, $k^{U}=0$ and $U=0$, so we see that indeed $\omega
_{0}=0$.

Also, it is easy to check that for a photon propagating along the horizon $%
l=0$. Indeed, if we write down the condition $k_{\mu }k^{\mu }=0$ on the
future horizon, we obtain that $k_{\phi }=0$. This agrees with previous
observations concerning the properties of trajectories on the horizon \cite%
{circkerr}, \cite{nh15}.

\section{Generalized Lemaitre frame}

\subsection{Form of metric}

It is instructive to reformulate the redshift value in the Lemaitre-like
coordinates $\rho ,\tau $. In contrast to the Kruskal ones, this frame is
based on free falling particles. The Lemaitre frame is well known for the
Schwarzschild metric. Now, we suggest its generalization valid for the
metric (\ref{met}).

The general theory of transformations that make the metric of a spherically
symmetric black hole regular, was developed in \cite{f}. For our goals, it
is sufficient to find a particular class of transformations that (i) makes the
metric regular on the horizon, (ii) generalizes the Lemaitre metric (in
particular, the metric should have $g_{\tau \tau }=-1$). We make the
transformation%
\begin{equation}
\rho =t+\int \frac{dr^{\ast }}{\sqrt{1-f}}\text{,}  \label{rof}
\end{equation}%
\begin{equation}
\tau =t+\int dr^{\ast }\sqrt{1-f}  \label{tauf}
\end{equation}%
where $r^{\ast }$ is given by (\ref{tc}). Eqs. (\ref{rof}), (\ref{tauf}) are
direct generalization of eqs. 102.1 of \cite{LL}. Then, it is easy to check
that

\begin{equation}
ds^{2}=-d\tau ^{2}+(1-f)d\rho ^{2}+r^{2}(\rho ,\tau )(d\theta ^{2}+\sin
^{2}\theta d\phi ^{2})\text{.}  \label{metL}
\end{equation}

On the horizon, $f=0$, the metric coefficient is regular, $g_{\rho \rho }=1$%
. In the particular case of the Schwarzschild metric, $f=1-\frac{r_{+}}{r}$
and we return to the standard formula for the Lemaitre metric, when $r$ is
expressed in terms of $\rho $ and $\tau $. The coordinates (\ref{rof}), (\ref%
{tauf}) are suitable for the description of a black hole including both the
outer R region and the contracting T$^{\_}$ one \cite{nov}. In a similar
way, one can use the expanding version that would result in a changing sign at 
$\tau $.

Now, we want to pay attention to some nice properties of the metric (\ref%
{metL}). The proper distance between points 1 and 2 calculated for a given $%
\tau $ is equal to $l=\int d\rho \sqrt{1-f}$. Requiring $d\tau =0$ in (\ref%
{tauf}) and substituting $dt$ into (\ref{rof}), we obtain from (\ref{tc}), (%
\ref{rof}) that%
\begin{equation}
l=r_{2}-r_{1}\text{.}  \label{dist}
\end{equation}

It is also instructive to calculate the velocity. Let, say, point 1 be fixed
and let us focus on the velocity of free fall $v=\frac{dl}{d\tau }$ of
a particle with $E=m$, where $r_{2}\equiv r$ changes depending on time.
Then, it is easy to  find from (\ref{mr}), (\ref{dist}) that%
\begin{equation}
v=-\sqrt{1-f}\text{.}
\end{equation}%
Taking the derivative ones more, we obtain $\frac{dv}{dr}=\frac{1}{2\sqrt{1-f%
}}\frac{df}{dr}$. On the horizon, this gives us%
\begin{equation}
\left( \frac{dv}{dr}\right) _{H}=\kappa \text{,}  \label{ka}
\end{equation}%
where we took into account that for our metric the surface gravity $\kappa =%
\frac{1}{2}\left( \frac{df}{dr}\right) _{H}$. The subscript "H" means that the
corresponding quantity is calculated on the horizon. Eq. (\ref{ka}) will be
used below. It is worth noting that for the extremal horizon ($\kappa =0$)
we have also $\left( \frac{dv}{dr}\right) _{H}=0$.

\subsection{Redshift: from Kruskal coordinates to Lemaitre ones}

The above frame is especially useful for the presentation of the redshift (%
\ref{12}). On the horizon, $f=0$. Then, in its vicinity, we obtain from (\ref%
{uU}), (\ref{rof}), (\ref{tauf}) that on the horizon%
\begin{equation}
v=\tau +C_{1}=\rho +C_{2}\text{,}
\end{equation}%
where $C_{1,2}$ are constants. As a result, we obtain from (\ref{12}) that%
\begin{equation}
\frac{\omega _{2}}{\omega _{1}}=\exp (\kappa (\tau _{1}-\tau _{2}))=\exp
(\kappa (\rho _{1}-\rho _{2}))\text{.}  \label{omk}
\end{equation}

Thus the Lemaitre frame allows us to present the resulting redshift along
the horizon in a simple and intuitively clear picture -- the redshift grows
(and, consequently, the emitter looks dimmer) exponentially with respect to
Lemaitre time that passes from emitting to observation.

In the last paragraph of Sec. II, we listed the general condition for the
geometrical optic to be valid. Now, we can express it in another way. Since
a physical wave packet has a finite length, parts of it will move away from
the black hole horizon even if its center is located exactly on the horizon.
Since the equation of light geodesics in the generalized Lemaitre frame
reads $dr/d\tau =1-\sqrt{1-f}$ for outward propagation, the Lemaitre time needed to leave the
vicinity of the horizon $r=r_{+}$ diverges as $\left\vert \ln ({r/r_{+}-1)}%
\right\vert $. Suppose, the emitter radiates light with the wavelength $%
\lambda $. Since in any case the wave packets cannot be smaller than $%
\lambda $, we can roughly estimate initial scale as $r-r_{+}\ \sim \lambda $%
. Then, we find that after the Lemaitre time $\tau /r_{+}\sim \ln {%
r_{+}/\lambda }\sim \ln {\omega _{0}r_{+}}$ the wave packet will reach the
scale of black hole horizon, the geometric optic approximation fails and, in
particular, Eq. (\ref{omk}) evidently breaks down.

\section{Photon emitted at the inner horizon}

Let us consider the situation similar to that considered above. An observer
moves beyond the event horizon $r_{+}$and approaches the inner horizon $%
r_{-}<r_{+}$. When he crosses it, he emits a photon. Another observer who
also crosses the inner horizon later, receives this very photon. What can be
said about its frequency?

\subsection{The coordinates and metric}

The metric between the outer and inner horizons represents so-called $T$
region \cite{nov}. For the definiteness, we consider $T^{-}$ region that
corresponds to a black hole but similar formulas are valid for the $T^{+}$
regions (white holes). Now, the metric can be formally obtained from (\ref%
{met}) if one takes into account that for $r_{\_}<r<r_{+}$the metric
function $f<0$, so spacelike and timelike coordinates mutually interchange.
We can write $f=-g$, 
\begin{equation}
y=t,T=-r\text{.}  \label{T}
\end{equation}%
Then, the metric can be rewritten in the form%
\begin{equation}
ds^{2}=-\frac{dT^{2}}{g}+gdy^{2}+T^{2}d\omega ^{2}\text{.}
\end{equation}

The equations of motion for a massive particle have the form%
\begin{equation}
m\dot{y}=\frac{P}{g}\text{,}
\end{equation}%
\begin{equation}
m\dot{T}=z\equiv \sqrt{P^{2}+m^{2}g}\text{,}  \label{z}
\end{equation}%
where $P=mu_{y}$ is the conserved momentum.

Now, the KS transformation somewhat changes and reads

\begin{equation}
U=\exp (-\kappa _{-}u)\text{,}  \label{ut}
\end{equation}%
\begin{equation}
V=\exp (\kappa _{-}v)\text{,}  \label{vt}
\end{equation}%
where $r^{\ast }$ is given by the formula

\begin{equation}
r^{\ast }=\int \frac{dr}{g}\text{.}  \label{tg}
\end{equation}%
The metric takes the form (\ref{mF}) with%
\begin{equation}
F=-\frac{g}{UV\kappa _{-}^{2}}\text{.}
\end{equation}%
where $\kappa _{-}$ is the surface gravity associated with the inner horizon
and $F\neq 0$ is finite there.

Repeating the calculations step by step, we arrive to the same formula (\ref{12})%
\begin{equation}
\frac{\omega _{2}}{\omega _{1}}=\frac{V_{1}}{V_{2}}.  \label{vin}
\end{equation}%
Now, we would like to pay attention that according to (\ref{uv}), $V$ is a
monotonically increasing function of $v$. It is seen from (\ref{uU}), (\ref%
{tc}) that for a fixed $u$, $\frac{\partial v}{\partial r}>0$. However, it
is seen from (\ref{T}) that event 2 that takes place after 1, has $%
r_{2}<r_{1}$. As a result, $v_{2}<v_{1}$ and $V_{2}<V_{1}$. Therefore, $%
\omega _{2}>\omega _{1}$ and now we have a blueshift. Thus this is related
to the fact that $r$ and $t$ coordinates change their character in the
region under discussion.

The results (\ref{omk}) and (\ref{vin}) can be united in one formula%
\begin{equation}
\frac{\omega _{2}}{\omega _{1}}=\exp [\left( \frac{dv}{dr}\right) _{H}(\rho
_{1}-\rho _{2})]=\exp [\left( \frac{dv}{dr}\right) _{H}(\tau _{1}-\tau _{2})]%
\text{,}
\end{equation}%
where $\tau _{2}>\tau _{1}$, $\rho _{2}>\rho _{1}$. For the outer horizon we
can use eq. (\ref{ka}) that gives us (\ref{omk}) and we have a redshift.
For the inner horizon, the counterpart of (\ref{ka}) gives us $\left( \frac{%
dv}{dr}\right) _{H}=-\kappa _{-}$, where now $\kappa _{-}=\frac{1}{2}%
\left\vert \frac{df}{dr}\right\vert _{H}$ is the surface gravity of the
inner horizon (where $\left( \frac{df}{dr}\right) _{H}<0$). As a result, we
obtain here a blueshift.

In a similar way, the procedure under discussion gives the same result when
an observer crosses the event horizon of a white hole moving outward from the $%
T^{+}$ to $R$ region. Then, he will detect all photons propagating along
this horizon to be blueshifted. In particular, this holds for the Schwarzschild
metric. Analogously, an observer entering $T^{+}$ region from the inner $R$
one (say, like in the Reissner-Nordstr\"{o}m metric)\ will see a redshift at
the inner horizon. In other words, in both situations (either black or white
hole) an observer crossing a horizon from the $T$ to $R$ region will see
a blueshift, while from the $R$ to $T$ region he will see a redshift.

\subsection{Relation to other effects}

In the previous subsection we have shown that the blueshift at the inner
horizon (and, consequently, the energy absorbed by the observer) grows
exponentially with the Lemaitre time between the moments of emission and
observation. Here we compare this interesting effect with others known in
literature.

If two particle collide, their energy $E_{c.m.}$ in the centre of mass frame
can be defined on the point of collision according to%
\begin{equation}
E_{c.m.}^{2}=-P_{\mu }P^{\mu }\text{,}
\end{equation}%
$P^{\mu }=p_{1}^{\mu }+p_{2}^{\mu }$ being the total momentum of two
particles. If particle 1 is massive and particle 2 is massless, $p_{1}^{\mu
}=mu^{\mu }$ and $p_{2}^{\mu }=k^{\mu }$, where we put the Planck constant
to unity. As a result,%
\begin{equation}
E_{c.m.}^{2}=m^{2}+2m\omega \text{.}
\end{equation}

In the example under discussion, if $V_{1}=O(1)$ and $V_{2}\rightarrow 0$,
the frequency $\omega _{2}\rightarrow \infty $ according to (\ref{vin}).
Then, $E_{c.m.}\rightarrow \infty $ as well and we encounter a counterpart
of the BSW effect near the inner horizon. But $V=0$ on the future horizon $%
U=0$ is nothing else than the bifurcation point \cite{inner} (see also below
for more details). Thus the present results for the blueshift agree with the
previous ones in the limit when the bifurcation point is reached.

There is also another issue to which we can compare the present
consideration. As is well known, near the inner (Cauchy) horizon an
instability develops inside black  holes. This happens when a decaying flux
of radiation coming from infinity crosses the event horizon and concentrates
near the inner one - see, e.g. Chapter 14.3.1 in \cite{fn}. (For a modern
review of the subject see \cite{ham}.) However, now we consider radiation
which is not coming from infinity but is emitted by an observer who crosses
the inner horizon. The resulting energy flux from an emitter at the inner
horizon appears to be finite, though it is not restricted from above if $%
V\rightarrow 0$.

Thus as far as the radiation near the inner horizon is concerned, we have
three situations: (i) the analogue of the BSW effect (relevant near the
bifurcation point), (ii) blueshift of a photon in the situation under
discussion (relevant near any point of the inner horizon, the blueshift is
in general finite), (iii) the instability of the inner horizon (infinite
blueshift due to concentration of radiation along the horizon). Cases (i)
and (ii) are closely related in the sense that in the limit when the point
where a photon is absorbed approaches the bifurcation point, one obtains (i)
from (ii). Meanwhile, in case (iii) the effect is unbounded and this 
points to a potential pathology connected with the nature of the \textit{inner}
 horizon.

\section{Special case: emission at the bifurcation point}

In the Sec. V, we discussed briefly such spacetimes that contain $T^{+}$
regions (white holes). Then the intersection between the future and past
horizons forms the so-called bifurcation point (sphere, if the angle
variables are taken into account), where it is possible to pass from the
white hole region to the black one. White holes and bifurcation points do
not arise in the situation when a black hole is formed due to gravitational
collapse and in this sense they are not feasible astrophysically. However, they
are inevitably present in the full picture of an eternal black-white hole.
Therefore, we consider such objects for theoretical reasons and for
completeness. In particular, in Sec. V, we saw that accounting for the
bifurcation point arises naturally in the connection between our problem and
the BSW effect. In doing so, it is a receiver that passes near the
bifurcation point.

In the present Section, we consider another case, when it is an emitter that
passes through this point at the moment of radiation. Consideration of the
frequency shift when a photon emitted from the bifurcation point is a
separate case that does not follow directly from the previous formulas. For
the Reissner-Nordstr\"{o}m-de Sitter metric, such a problem was considered
in Sec. IV b of \cite{lake}. It follows from the corresponding results that
different cases are possible here: $\omega _{2}<\omega _{1}$, $\omega
_{2}=\omega _{1}$, $\omega _{2}>\omega _{1}.$ On the first glance, this
disagrees with our results described above since we obtained either redshift
(for the event horizon of a black hole or inner horizon of a white hole) or
blueshift (for the inner one in a black hole or event horizon of a white
hole). Fortunately, this contradiction is illusory. Now, we will explain how
one can obtain the results for the bifurcation point from ours. To this end,
we compare (i) the generic situation and (ii) that with crossing the
bifurcation point and trace how (ii) arises from (i) within the limiting
transition.

For our purposes, it is sufficient to discuss the simplest metric that
possesses the bifurcation point, so we can imply it to be, say, the
Schwarzschild one. We assume that the emitter 1 moves from the inner
expanding $T^{+}$ region (i.e. white hole) \cite{nov}, crosses the past
horizon and enters the R region. Afterwards, it crosses the event horizon
falling into a black hole. Let, as before, the emitter and receiver have
equal masses $m_{1}=m_{2}=m$. However, now we cannot put \ $E_{1}=E_{2}.$
This is because a particle with $E=m$ would escape to infinity instead of
falling into a black hole. We remind the reader that up to now, in all our
considerations an emitter and an observer are set to be at rest in infinity.
However, the most gereral case can easily be obtained by adding
corresponding Lorentz boosts. In the present subsection we meet the
situation where this procedure is needed.

Therefore, we must use more general formula based on (\ref{omf})%
\begin{equation}
\frac{\omega _{2}}{\omega _{1}}=\frac{E_{2}}{E_{1}}\frac{V_{1}}{V_{2}}\text{.%
}  \label{gen}
\end{equation}%
The first factor can be interpreted as a Lorentz boost responsible for the
Doppler effect. For $E_{1}=E_{2}$ we return to the case considered by us
above but now the first factor is not equal to one and plays now a crucial
role.

If, by assumption, particle 1 falls into a black hole, this means that it
must bounce from the potential barrier in the turning point $r=r_{0}$.
According to equations of motion (\ref{mr}), this means that%
\begin{equation}
E=m\sqrt{f(r_{0})}\text{.}  \label{ef}
\end{equation}

If $r_{0}\rightarrow r_{+}$, $f(r_{0})\rightarrow 0$, so $E\rightarrow 0$ as
well. More precisely, it is seen from (\ref{kar}), (\ref{uvc}) that%
\begin{equation}
E\sim \sqrt{r_{0}-r_{+}}\sim \sqrt{\left\vert U\right\vert V}\text{.}
\end{equation}%
As a result,%
\begin{equation}
\frac{\omega _{2}}{\omega _{1}}\sim \frac{\alpha E_{2}}{V_{2}}\text{, }%
\alpha \equiv \sqrt{\frac{\left\vert V\right\vert _{1}}{U_{1}}}\text{.}
\end{equation}

In the limit when the trajectory of particle 1 passes closer and closer to
the bifurcaiton point $U=0=V$, $\alpha $ remains finite. Using equations of
motion in the T region (see the previous section), it is easy to show that
in the limit $V\rightarrow 0$, $U\rightarrow 0$, the component of the
velocity $u^{U}$ contains just this factor $\alpha $.

Thus depending on relation between $\alpha $ and $V_{2}$ one can obtain any
result for $\omega _{2}$ (redshift, blueshift, the absence of the frequency
shift). In this sense, the general formula (\ref{gen}) reproduces both
"standard" fall of the emitter in a black hole and the behavior of the
emitter that passes through the bifurcation point.

\section{Extremal horizon}

Let an observer crosses the (ultra) extremal horizon $r_{+}$. By definition,
this means that near it the metric function is
\begin{equation}
f\sim (r-r_{+})^{n}
\end{equation}%
where $n=2$ in the extremal case and $n=3,4...$ in the ultraextremal one.
The difference with the nonextremal case consists in a different nature of
transformation making the metric regular. Let the two-dimensional part of
the  metric has the same form as in (\ref{met}). The subsequent procedure
is known  - see, e.g., \cite{lib}, \cite{bron} (Sec. 3.5.1). We use the
same coordinates $u$, $v$ and want to find appropriate coordinates $U,$ $V$,%
\begin{equation}
V=V(v)\text{, }U=U(u)\text{.}
\end{equation}

Now, we are interested in the situation with emission of a photon exactly
along the horizon.Then, near the horizon it follows for the tortoise
coordinate (\ref{tc}) that%
\begin{equation}
r-r_{+}\sim \left\vert r^{\ast }\right\vert ^{\frac{1}{1-n}}\text{,}
\end{equation}%
\begin{equation}
f\sim \left\vert r^{\ast }\right\vert ^{\frac{n}{1-n}}\text{.}
\end{equation}%
We consider the metric near the future horizon where $v$ is finite, $r^{\ast
}\rightarrow -\infty $, $u=v-2r^{\ast }\rightarrow +\infty $. We have%
\begin{equation}
f\sim u^{\frac{n}{1-n}}\text{.}  \label{fu}
\end{equation}

We try a transformation that behaves like 
\begin{equation}
U\sim u^{-\frac{1}{n-1}}\text{,}  \label{uextr}
\end{equation}%
so that $U\rightarrow 0$. Then, it is easy to check that the metric has the
form (\ref{mF}) where $F\neq 0$ is finite on the horizon. To find the
frequency, we must use the expression for $u^{U}$ (\ref{mu}) in which now (%
\ref{uextr}) is valid, so $\frac{dU}{du}\sim u^{\frac{n}{1-n}}$. \ It is
seen from (\ref{fu}) that $u^{U}\rightarrow const$ on the horizon and it does
not contain $V$. Taking into account that $k^{V}$ is a constant along the
horizon generator as before, we come to the conclusion that $V$ drops out and $%
\frac{\omega _{2}}{\omega _{1}}=const.$ We see that in the horizon limit the
quantity $V$ does not enter the frequency. In this sense, $\frac{\omega _{2}%
}{\omega _{1}}$ does not change along the horizon, so redshift or blueshift
is absent.

In a sense, it is quite natural. Indeed, the extremal horizon is the double
one. The inner and outer horizons merge. But for an inner horizon we had
a blueshift, for the outer one we had a redshift. Together, they mutually cancel
and produce no effect.

The absence of the redshift or blueshift formally agrees with (\ref{omk}) if
one puts $\kappa =0$ there. However, for (ultra)extremal black holes the
Kruskal-like transformation looks very different, so we could not use eq. (%
\ref{omk}) directly. Therefore, it was not obvious in advance, whether or
not the redshift for the extremal horizon can be obtained as the extremal
limit of a nonextremal one. Now, we see that this is the case.

\section{Summary}

Thus we showed that for emission along the outer horizon redshift occurs and
we derived a simple formula that generalized the one previously found in
literature. We also showed that along the inner horizon blueshift occurs and
found its relation with the BSW effect. We also showed how the previously known
results for the emission at the bifurcation point are reproduced from a
general formula and lead to a diversity of situations (redshift, blueshift
or the absence of frequency shift). For (ultra)extremal horizons the effect
is absent.

These observations have a quite general character in agreement with the
universality of black hole physics. We also generalized the Lemaitre frame
and in this frame derived a simple and instructive formula for a redshift
along the horizon in terms of the Lemaitre time and the surface gravity.

\section*{Acknowledgements}

The work was supported by the Russian Government Program of Competitive
Growth of Kazan Federal University.

\end{document}